\documentclass[aps,pra,superscriptaddress,amsmath,amssymb,preprintnumbers,showpacs,twocolumn]{revtex4-1}

\usepackage{amssymb}
\usepackage{graphicx}
\usepackage{dcolumn}
\usepackage{bm}
\usepackage{color}

\begin{document}

\newcommand{\beq}{\begin{equation}}
\newcommand{\eeq}{\end{equation}}
\newcommand{\barr}{\begin{eqnarray}}
\newcommand{\earr}{\end{eqnarray}}

\def\bra#1{\langle{#1}|}
\def\ket#1{|{#1}\rangle}
\def\sinc{\mathop{\text{sinc}}\nolimits}
\def\cV{\mathcal{V}}
\def\cH{\mathcal{H}}
\def\cT{\mathcal{T}}
\renewcommand{\Re}{\mathop{\text{Re}}\nolimits}
\newcommand{\tr}{\mathop{\text{Tr}}\nolimits}

\definecolor{dgreen}{rgb}{0,0.5,0}
\newcommand{\dgreen}{\color{dgreen}}
\newcommand{\red}{\color{red}}
\newcommand{\green}{\color{dgreen}}
\newcommand{\blue}{\color{blue}}
\newcommand{\magenta}{\color{magenta}}
\newcommand{\cyan}{\color{cyan}}
\newcommand{\yellow}{\color{yellow}}
\newcommand{\orange}{\color[named]{Orange}}
\newcommand{\mahogany}{\color[named]{Mahogany}}
\newcommand{\purple}{\color[named]{Purple}}
\newcommand{\gray}{\color[named]{Gray}}
\newcommand{\black}{\color{black}}
\definecolor{delete}{cmyk}{0.5,0,0,0}

\newcommand{\CMT}[1]{\textbf{\color{red}[[[#1]]]}}
\newcommand{\REV}[1]{\textbf{\color{red}[[#1]]}}
\newcommand{\RED}[1]{{\color{red}#1}}
\newcommand{\BLUE}[1]{{\color{blue}#1}}
\newcommand{\GREEN}[1]{{\green #1}}
\newcommand{\MAGENTA}[1]{{\magenta #1}}
\newcommand{\CYAN}[1]{{\cyan #1}}
\newcommand{\DEL}[1]{}

\def\Nhalf{{\scriptstyle{\scriptstyle N\over\scriptstyle2}}}

\title{Photon distribution at the output of a beam splitter for imbalanced input states}

\author{Hiromichi Nakazato}
\affiliation{Department of Physics, Waseda University, Tokyo 169-8555, Japan}

\author{Saverio Pascazio}
\affiliation{Dipartimento di Fisica and MECENAS, Universit\`a di Bari, I-70126 Bari, Italy}
\affiliation{INFN, Sezione di Bari, I-70126 Bari, Italy}

\author{Magdalena Stobi\'nska}
\affiliation{Institute of Theoretical Physics and Astrophysics,
University of Gda\'nsk, ul. Wita Stwosza 57, 80-952 Gda\'nsk, Poland\\
\& National Quantum Information Center of Gda\'nsk, 81-824 Sopot, Poland}
\affiliation{Institute of Physics, Polish Academy of Sciences, Al. Lotnik\'ow 32/46, 02-668 Warsaw, Poland}

\author{Kazuya Yuasa}
\affiliation{Department of Physics, Waseda University, Tokyo 169-8555, Japan}

\begin{abstract}
In the Hong--Ou--Mandel interferometric scheme, two identical photons that illuminate a balanced beam splitter always leave through the same exit port. Similar effects have been predicted and (partially) experimentally confirmed for multi-photon Fock-number states. In the limit of large photon numbers, the output distribution follows a $(1-x^2)^{-1/2}$ law, where $x$ is the normalized imbalance in the output photon numbers at the two output ports. We derive an analytical formula that is also valid for imbalanced input photon numbers with a large total number of photons, and focus on the extent to which the hypothesis of perfect balanced input can be relaxed, discussing the robustness and universal features of the output distribution.
\end{abstract}
\maketitle

\section{Introduction}
\label{sec:intro}
Two identical photons, impinging on a balanced beam splitter, always leave through the same exit port, due to the Hong--Ou--Mandel (HOM) interference \cite{Hong-Ou-Mandel,shih}. 
Similar effects can be observed for multi-photon Fock-number states: photons will leave the beam splitter only in certain configurations, for example such that the difference between the occupations of the exit ports is even, while an odd difference never occurs \cite{CST,Stobinska12,Stobinska15}. These results have been partially experimentally confirmed for photons \cite{Spasibko14}, although the existence of the odd-even structure was not demonstrated. 
Similar effects have been discussed for atomic Bose-Einstein condensates \cite{Bouyer97,Lucke11}, in terms of spin dynamics, modeled by the population imbalance.

In this Article we shall investigate the photon distribution at the output ports of a balanced beam splitter when the input state is a product of number states. If the the numbers of photons at the two input ports are perfectly balanced, the output distribution follows a $(1-x^2)^{-1/2}$ law, where $x$ is the normalized imbalance in the output photon numbers at the two output ports [see (\ref{balanced_as}) in the following].  However, it is interesting to ask what happens when the input photon state is not perfectly balanced. This is relevant because of practical reasons, as photon numbers may fluctuate, say according to a Poisson distribution, but also in view of future possible applications. We shall prove that the output distribution is robust, and some of its features remain unchanged, even if the hypothesis of perfectly balanced input is relaxed. In fact, we will focus on the extent to which such hypothesis can be relaxed.

Our interest in these phenomena is threefold. On one hand, they offer perspectives in applications, as the output distribution can be viewed as a generalized NOON state \cite{noon}, in the sense that photons bunch and tend to exit the beam splitter at only one of its output ports. These states have remarkable applications in metrology \cite{metrology}, as they lead to the Heisenberg limit.
Also, the general features that emerge from our analysis are reminiscent of typical behavior \cite{YI,typbec,FNPPSY} in optics and cold atomic physics \cite{molmer,SBRK,CD}, bearing consequences on the foundations of statistical mechanics \cite{Tasaki,Winter,Popescu1}.
Finally, there are remarkable similarities with the physics of continuous-time quantum walks, where rigorous results have been obtained \cite{Konno1,Konno2}.

The main result of this Article will be the evaluation of the photon distribution at the output ports of a beam splitter, when the total number of impinging photons is large and imbalanced. We will formulate the problem exactly and then display its asymptotic features.
In Sec.\ \ref{sec:bs} we introduce notation and set up the mathematical description of a beam splitter. 
The balanced input case is solved in Sec.\ \ref{sec:balanced}, while the imbalanced input case is solved in Sec.\ 
\ref{sec:imbalanced}\@.
The universal features that emerge in the latter case are discussed in Sec.\ \ref{sec:comments}, where the 
(average) output distribution is shown to follow a $(1-x^2)^{-1/2}$ law, $x$ being the normalized imbalance in the output photon numbers at the two output ports. On average, this law is \emph{robust}, namely insensitive to the input imbalance (the upper limit to the fluctuations being Poissonian). The statistical fluctuations are further analyzed in 
Sec.\ \ref{sec:twopoints}, where the characteristics of the two-body correlation function of the probability distribution are computed. We conclude in Sec.\ \ref{sec:concl} by discussing further perspectives and possible applications.

\section{Beam splitter}
\label{sec:bs}
\begin{figure}
\centering
\includegraphics[width=0.4\textwidth]{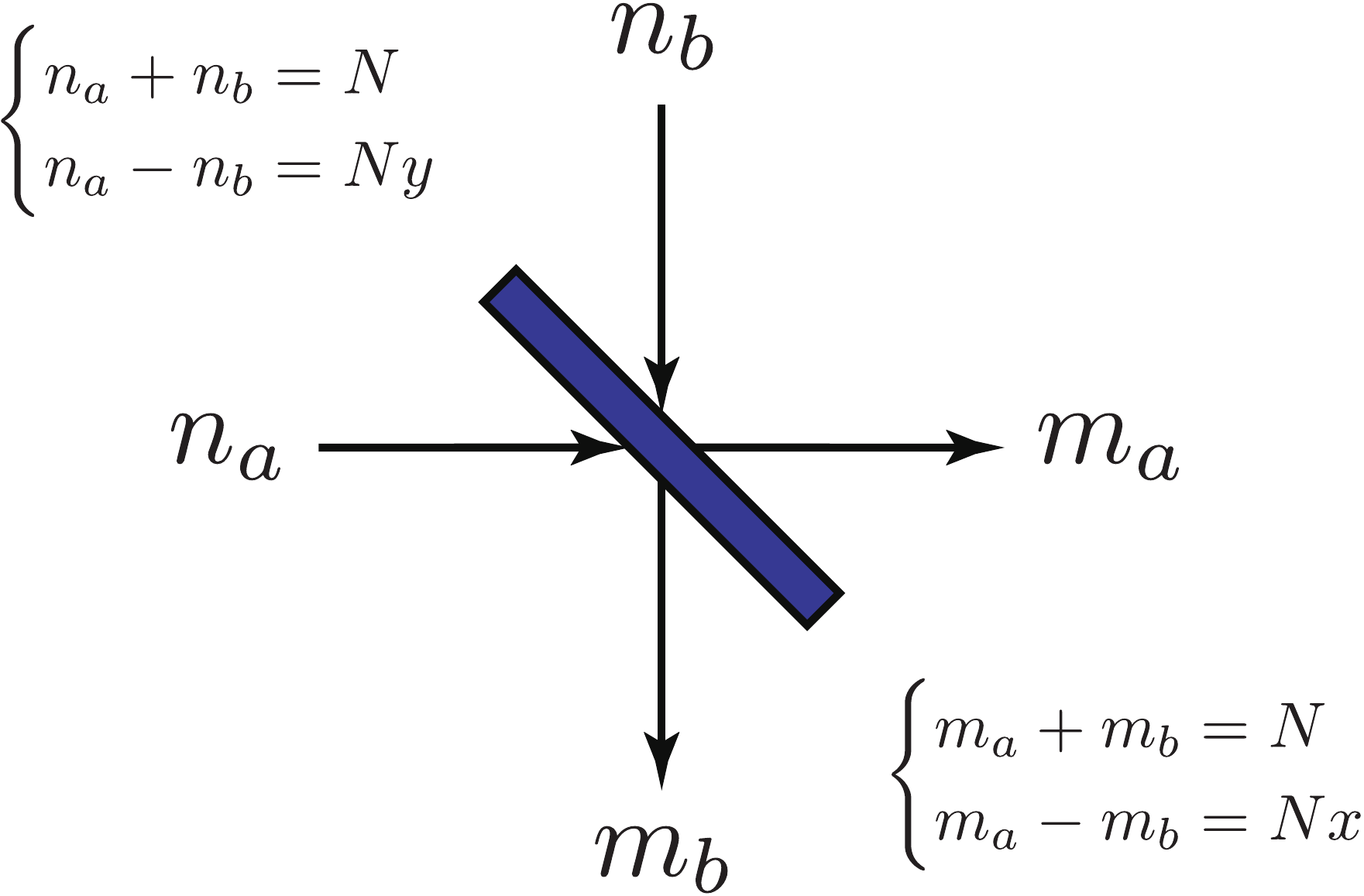}
\caption{A beam splitter: $n_a$ and $n_b$ photons illuminate ports $a$ and $b$, respectively, and the total number of photons is fixed $n_a+n_b=N$; $m_a$ and $m_b$ photons exit through ports $a$ and $b$, respectively.
The input and output imbalances read $Ny=n_a-n_b$ and $Nx=m_a-m_b$, respectively.}
\label{fig:setup}
\end{figure}

Consider the beam splitter in Fig.\ \ref{fig:setup}, where $n_a$ and $n_b$ photons illuminate ports $a$ and $b$, respectively. Let the total number of photons be fixed $n_a+n_b=N$, and the input state be given by $\ket{n_a,n_b}=\ket{n_a,N-n_a}$. We intend to study the photon distribution at the output ports, namely the 
amplitude of having $m_a$ and $m_b$ photons at output ports $a$ and $b$, respectively.
Since the beam splitter preserves the total number of photons, the output photon numbers $m_a$ and $m_b$ are also constrained as $m_a+m_b=N$.

We are interested in the large-$N$ limit, but let us start by recalling what happens in the simplest case $(n_a,n_b)=(1,1)$. Then, the output is either $(m_a,m_b)=(2,0)$ or $(0,2)$. Only the two extreme cases appear, while the balanced output $(m_a,m_b)=(1,1)$ is suppressed. This is the HOM interference \cite{Hong-Ou-Mandel,shih}, due to photon bunching.
If the input photon number $N$ is greater than $2$, the two-peak structure in the probability distribution is blurred, but a similar structure remains in the large-$N$ limit. Moreover, such a structure will be shown to be very robust against the fluctuations in the imbalance in the input photon numbers. 

The action of the beam splitter is described by the unitary operator
\begin{equation}
\hat{U} = e^{-\xi(\hat{a}^\dagger \hat{b}-\hat{b}^\dagger a)}=e^{\hat{J}_-\tan\xi}e^{\hat{J}_3\ln\cos\xi}e^{-\hat{J}_+\tan\xi},
\end{equation}
where $\xi=\pi/4$ for a 50:50 beam splitter, $\hat{J}_3=\hat{a}^\dagger\hat{a}-\hat{b}^\dagger \hat{b}$, $\hat{J}_+= \hat{a}^\dagger \hat{b}$, and $\hat{J}_-= \hat{b}^\dagger \hat{a}=\hat{J}_+^\dagger$ \cite{perelomov}, with $\hat{a}$ and $\hat{b}$ being the canonical annihilation operators of photons in the two modes. 
The input state $\ket{n_a,N-n_a}$ is obtained 
from the (normalized) state $|0,N\rangle$ by \cite{perelomov,sakurai1994modern}
\begin{equation}
\ket{n_a,N-n_a}=\sqrt{{(N-n_a)!\over n_a!\,N!}}(\hat{J}_+)^{n_a}|0,N\rangle.
\end{equation}
The amplitude to get output $\ket{m_a,N-m_a}$ reads
\begin{widetext}
\begin{align}
\langle m_a,N-m_a|\hat{U}|n_a,N-n_a\rangle
&={1\over N!}\sqrt{(N-m_a)!\,(N-n_a)!\over m_a!\,n_a!}(\cos\xi)^{2m_a-N}\langle0,N|(\hat{J}_-)^{m_a}e^{\hat{J}_-\sin\xi\cos\xi}e^{-\hat{J}_+\tan\xi}(\hat{J}_+)^{n_a}|0,N\rangle\nonumber\\
&=\sqrt{(N-m_a)!\over m_a!}{(\cos\xi)^{2m_a-N}\over\sqrt{n_a!\,(N-n_a)!}}
\left.
\left({\partial\over\partial\alpha}\right)^{m_a}[\alpha^{n_a}(1+\alpha\beta)^{N-n_a}]
\right|_{\alpha=\sin\xi\cos\xi,\beta=-\tan\xi}\nonumber\\
&\equiv A_N(x,y) \qquad (Nx=m_a-m_b=2m_a-N,\,Ny=n_a-n_b=2n_a-N),
\vphantom{\sqrt{(N-m_a)!\over m_a!}}
\label{eq:amp}
\end{align}
\end{widetext}
where we have introduced the normalized imbalances $y$ and $x$ in the input and output photon numbers, respectively, both ranging in $-1\le x,y\le1$.
This is our starting formula. 

\begin{figure}
\centering
\includegraphics[width=0.48\textwidth]{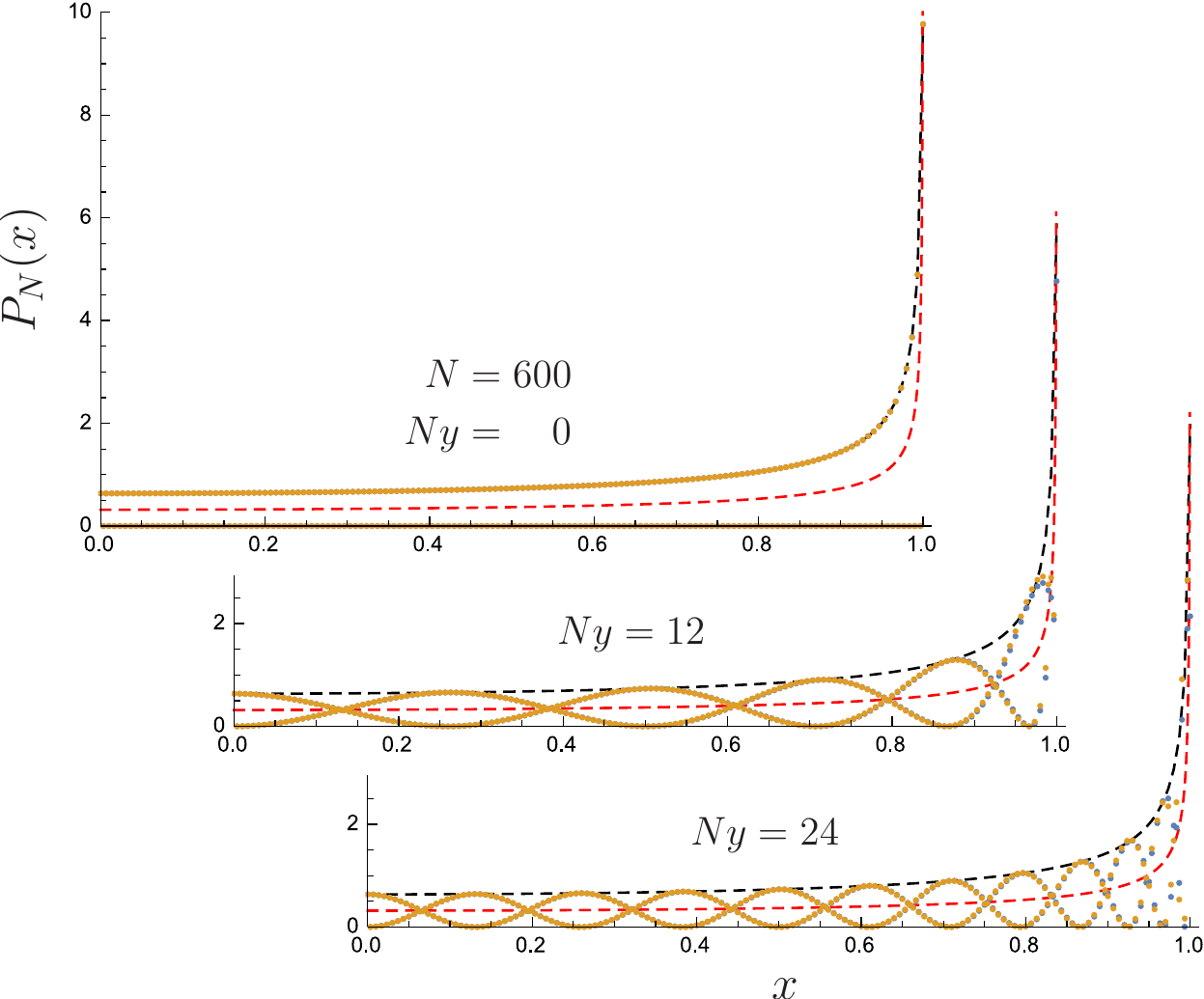}
\caption{(Color online) Output distributions $P_N(x)$ in Eq.\ (\ref{eqn:Px}), based on the approximate formula (\ref{imbalanced}) (orange points) and exact numerical evaluation (blue points), with $N=600$ for different input imbalances $Ny=0,12,24$. 
Note that $\sqrt{N}=\sqrt{600}\simeq24.5$.
All distributions are symmetric in $x$ and are plotted only for $x\ge0$. Since $N=600$ is even, only even output imbalances $Nx$ are allowed, and $P_N(x)$ vanishes for $Nx=0,\pm4,\pm8,\ldots$ when $y=0$.
In all panels, the upper (black) dashed curve is the upper envelope of $P_N(x)$ for the balanced input case $y=0$ based on Eq.\ (\ref{balanced}), and the lower (red) dashed curve is $P(x)=(1/\pi)(1-x^2)^{-1/2}$ given in Eq.\ (\ref{eqn:Pave}). 
Since the approximation is very good, the discrepancy between the approximate formula (orange points) and the exact numerical evaluation (blue points) is invisible except for $|x|\sim1$.
}
\label{fig:Px}
\end{figure}

\section{Balanced photon input $\bm{y=0}$}
\label{sec:balanced}
We first consider the balanced-input case $y=0$. This implies that the total photon number $N$ is even, and only even output imbalances $Nx$ are allowed.
The evaluation of the last factor yields $[m_a=(N/2)(1+x)]$
\begin{align}
&
\left. \left({\partial\over\partial\alpha}\right)^{{N\over2}(1+x)}[\alpha^{{N\over2}}(1+\alpha\beta)^{{N\over2}}]\right|_{\alpha=1/2,\beta=-1} \nonumber\displaybreak[0]\\
&={[{N\over2}(1+x)]!\over2\pi i}\oint dz\,{z^{{N\over2}}(1-z)^{{N\over2}}\over(z-{1\over2})^{{N\over2}(1+x)+1}}\nonumber\displaybreak[0]\\
&={[\Nhalf(1+x)]!\over2\pi}\left({1\over2}\right)^{{N\over2}(1-x)}\oint d\theta\,e^{-i{N\over2}(1+x)\theta}(1-e^{2i\theta})^{{N\over2}}\nonumber\displaybreak[0]\\
&=(-1)^{{N\over4}(1+x)}[\Nhalf(1+x)]!\left({1\over2}\right)^{{N\over2}(1-x)}
\begin{pmatrix}{N\over2}\\{N\over4}(1+x)\end{pmatrix},
\end{align}
where the quantity ${N\over4}(1+x)$ is assumed to be integer, otherwise we get a null result.
Therefore, the amplitude is found to be expressed analytically as
\begin{equation}
A_N(x,0) 
 =(-1)^{{N\over4}(1+x)}{\sqrt{[{N\over2}(1+x)]!\,[{N\over2}(1-x)]!}\over2^{{N\over2}}[{N\over4}(1+x)]!\,[{N\over4}(1-x)]!}
\label{balanced}
\end{equation}
for integer $\frac{m_a}{2}={N\over4}(1+x)$, otherwise $A_N(x,0)=0$. This formula is exact and coincides with the result obtained in Ref.\ \cite{CST}, where an analogy is drawn with the vector model \cite{vector}.
Since the amplitude identically vanishes every two (``even'') points, the probability distribution appears as a rapidly oscillating function of $x$. 
Observe that the odd and even ``branches" of (\ref{balanced}) ``compete" at the edges $|x|=1$ of the distribution, yielding wild oscillations. 
See the upper panel in Fig.\ \ref{fig:Px}, where the distribution 
\begin{equation}
P_N(x)=\frac{N}{2}|A_N(x,y)|^2
\label{eqn:Px}
\end{equation}
is plotted for $N=600$ and $y=0$. This distribution has a comb-like structure, oscillating between its local maxima
[square of Eq.\ (\ref{balanced})] and $0$.
We will come back to this observation when we will consider the imbalanced-input case with $y\neq0$ [see (\ref{13})].

The asymptotic behavior of $A_N(x,0) $ for large $N$ is easily evaluated by using the Stirling formula, 
\begin{equation}
A_N(x,0) \sim (-1)^{{N\over4}(1+x)}{2\over\sqrt{\pi N}(1-x^2)^{{1\over4}}}.
\label{balanced_as}
\end{equation}
The average between the upper and lower envelopes of $P_N(x)$ in the upper panel of Fig.\ \ref{fig:Px} for $y=0$ is just half of the upper envelope,
\begin{equation}
P(x)
=\frac{1}{\pi\sqrt{1-x^2}},
\label{eqn:Pave}
\end{equation}
which is normalized $\int_{-1}^1P(x)dx=1$, and is plotted in Fig.\ \ref{fig:Px} (dashed line).

\section{Imbalanced photon input $\bm{y\neq0}$}
\label{sec:imbalanced}
The evaluation of (\ref{eq:amp}) for nonvanishing $y$ is more involved and requires the calculation of the last factor in (\ref{eq:amp}). Let us first focus on this factor and rewrite it as
\begin{multline}
\left.
 \left({\partial\over\partial\alpha}\right)^{{N\over2}(1+x)}[\alpha^{{N\over2}(1+y)}(1+\alpha\beta)^{{N\over2}(1-y)}]\right|_{\alpha=1/2,\beta=-1} \\
= {[{N\over2}(1+x)]!\over2\pi}2^{-{N\over2}(1-x)}i^{-{N\over2}(1-y)}I_{{N\over2}y},
\end{multline}
where
\begin{equation}
I_n\equiv2^N\oint d\theta\left(\sin{\theta\over2}\right)^{{N\over2}-n}\left(\cos{\theta\over2}\right)^{{N\over2}+n}e^{-i{N\over2}x\theta},
\end{equation}
with $n=Ny/2$.
It is not difficult to derive the recursion relation 
\begin{equation}
I_n={{N\over2}-n-1\over{N\over2}+n+1}I_{n+2}-{iNx\over{N\over2}+n+1}I_{n+1}.
\label{exactrecur}
\end{equation}

\subsection{Sub-Poissonian case: $\bm{n=o(\sqrt{N})$}}
\label{sec:subp}

Equation (\ref{exactrecur}) is exact. 
For $n\ll N$, $n$ in the coefficients can be neglected altogether and Eq.\ (\ref{exactrecur}) reduces to 
\begin{equation}
I_n^{(0)}\sim I_{n+2}^{(0)}-2ixI_{n+1}^{(0)}.
\end{equation}
[As we shall see in the following subsection, this amounts to requiring $n=o(\sqrt{N})$, namely sub-Poissonian imbalance.]
The solution to this approximate recursion relation is easily found and yields an explicit expression for $I_n^{(0)}$ as a function of the two initial terms $I_0$ and $I_1$,
\begin{equation}
I_n^{(0)}={p^n-q^n\over p-q}I_1-{pq(p^{n-1}-q^{n-1})\over p-q}I_0.
\label{recur}
\end{equation}
The two parameters $p$ and $q$ are given by
\begin{equation}
p,\,q=ix\pm\sqrt{1-x^2}=\pm e^{\pm i\tan^{-1}{x\over\sqrt{1-x^2}}},
\end{equation}
so that the function $I_n^{(0)}$ is found to be approximately given, for small $n\ll N$, by
\begin{align}
I_n^{(0)} \sim{}& i(-i)^n{\sin [n({\pi\over2}+\tan^{-1}{x\over\sqrt{1-x^2}})]\over\sqrt{1-x^2}}I_1
\nonumber \\
&{}+ i(-i)^{n-1}{\sin [(n-1)({\pi\over2}+ \tan^{-1}{x\over\sqrt{1-x^2}})]\over\sqrt{1-x^2}}I_0.
\label{recur2} 
\end{align}
The term $I_0$ is essentially the same as in the balanced-input case,
\begin{equation}
I_0=2\pi i^{{N\over2}}(-1)^{{N\over4}(1+x)}\begin{pmatrix}{N\over2}\\{N\over4}(1+x)\end{pmatrix}_0,
\end{equation}
where the subscript $_0$ signifies that the lower entry in the binomial is an integer, otherwise the term vanishes.
The calculation of $I_1$ is a bit involved but can be done explicitly.
We rewrite the relevant integral in the following way
\begin{multline}
\oint d\theta\, e^{-i{N\over2}(1+x)\theta}(1-e^{2i\theta})^{{N\over2}}{1+e^{i\theta}\over1-e^{i\theta}}
\\
=i(-2i)^{{N\over2}}\oint d\theta\,\Bigl[(\sin\theta)^{{N\over2}-1}+ix(\sin\theta)^{{N\over2}}\Bigr]\,e^{-i{Nx\over2}\theta},
\end{multline}
which is easily integrated, yielding
\begin{multline}
I_1=2\pi i^{1-{N\over2}}\,\biggl[2(-1)^{{N\over4}(1-x)-{1\over2}}\begin{pmatrix}{N\over2}-1\\{N\over4}(1+x)-{1\over2}\end{pmatrix}_0\\
{} + x(-1)^{{N\over4}(1-x)}\begin{pmatrix}{N\over2}\\{N\over4}(1+x)\end{pmatrix}_0\biggr].
\end{multline}
Let us postpone the corresponding solution for the amplitude $A_N$ to the following subsection.

\subsection{Poissonian case: $\bm{n=O(\sqrt{N})}$}
\label{sec:poiss}

The above estimation (\ref{recur2}) is valid only when the corrections of order $n/N$ do not accumulate to give a correction of order 1.
Since there are $n$ factors, each of which contributes a correction of order $n/N$ to $I_n$, the approximation is valid for $n=o(\sqrt{N})$.
However, when $n=O(\sqrt{N})$, one needs to take these contributions into account. This can be achieved by plugging the ansatz 
\begin{equation}
I_n=I_n^{(0)} e^{\frac{f_n}{N}}
\label{inin0}
\end{equation}
into (\ref{exactrecur}), and by expanding the recursive formula in $n/N$. One gets
\begin{equation}
f_{n+1} \simeq f_0 + n(n+1) \quad \longrightarrow \quad f_n \simeq n^2.
\label{iterf}
\end{equation}
so that the solution in (\ref{recur2}) must be simply multiplied by the factor $e^{n^2/N}=e^{Ny^2/4}$. This factor is crucial when one deals with the Poissonian case, while it can be neglected when $n=o(\sqrt{N})$.
Putting everything together, we finally arrive at the analytic expression for the amplitude
\begin{widetext}
\begin{align}
A_N(x,y) 
&\sim-{1\over2^{{N\over2}}}\sqrt{[{N\over2}(1+x)]!\,[{N\over2}(1-x)]!\over[{N\over2}(1+y)]!\,[{N\over2}(1-y)]!}e^{{N\over4}y^2} \nonumber\\
&\qquad\times\left\{{\sin[{Ny\over2}({\pi\over2}+\tan^{-1}{x\over\sqrt{1-x^2}})]\over\sqrt{1-x^2}}\left[2(-1)^{-{N\over4}(1+x)-{1\over2}}
\begin{pmatrix}{N\over2}-1\\{N\over4}(1+x)-{1\over2}\end{pmatrix}_0
+x(-1)^{{N\over4}(1+x)}\begin{pmatrix}{N\over2}\\{N\over4}(1+x)\end{pmatrix}_0\right]\right.\nonumber\\
&\qquad\qquad\left.
{}+{\sin[({Ny\over2}-1)({\pi\over2}+\tan^{-1}{x\over\sqrt{1-x^2}})]\over\sqrt{1-x^2}}(-1)^{{N\over4}(1+x)}
\begin{pmatrix}{N\over2}\\{N\over4}(1+x)\end{pmatrix}_0\right\},
\label{imbalanced}
\end{align}
\end{widetext}
where the subscript $_0$ signifies that the lower entry in the binomial [be it ${N\over4}(1+x)-{1\over2}$ or
${N\over4}(1+x)$] is an integer, otherwise the term vanishes.
This expression is one of our main results: it is valid for $0\le Ny\ll N$ and reduces to the previous result (\ref{balanced}) when $y=0$. (Incidentally, we notice that only the condition $0\le Ny\ll N$ is required, so that in practice $N$ need not be very large.)
Observe the presence of a nontrivial $x$ dependence appearing in the sinusoidal function once the input imbalance has been incorporated.
Roughly speaking, one expects that about $Ny/2$ oscillations appear in the probability distribution.
For negative input imbalance $-N\ll Ny<0$, a similar expression is obtained, with the variable $y$ replaced by $|y|$ and multiplied by a phase factor $(-1)^{{N\over2}(1+x)}$ [see (\ref{eq:amp}) with $\xi=\pi/4$]. 

The corresponding distribution $P_N(x)$ defined in (\ref{eqn:Px}) is plotted in Fig.\ \ref{fig:Px}, for $N=600$ and the input imbalances $Ny=12$ and $24$. 
Note that $\sqrt{N}=\sqrt{600}\simeq24.5$.
The agreement is excellent, as one starts to observe deviations only for $|x|\sim 1$.
The distribution $P_N(x)$ displays again rapid (point by point) oscillations, but one notices the presence of two slowly oscillating envelopes, that are obtained if one separately joins points for integer $\frac{N}{4}(1+x)+\frac{1}{2}$ and points for integer $\frac{N}{4}(1+x)$.

For large $N$, the amplitude is approximated by the following function [apart from the total phase $(-1)^{{N\over2}(1+x)}$ for negative $y$],
\begin{widetext}
\begin{align}
A_N(x,y)
={}&{-}{2\over\sqrt{\pi N}}{e^{{N\over4}y^2}\over(1+y)^{{N\over4}(1+y)}(1-y)^{{N\over4}(1-y)}(1-y^2)^{{1\over4}}}\nonumber\displaybreak[0]\\
&{}\times\left((-1)^{{N\over4}(1+x)+{1\over2}}\Big\vert_0{\sin[{N|y|\over2}({\pi\over2}+\sin^{-1}x)]\over\sqrt{1-x^2}}(1-x^2)^{{1\over4}}
-(-1)^{{N\over4}(1+x)}\Big\vert_0\cos[\tfrac{N|y|}{2}(\tfrac{\pi}{2}+\sin^{-1}x)]{1\over(1-x^2)^{{1\over4}}}\right),
\label{13}
\end{align}
\end{widetext}
where the subscript $_0$ signifies that the exponent of $(-1)$ is an integer, otherwise the term preceding the vertical bar vanishes.
The expression (\ref{13}) is our second main result, being a consequence of (\ref{imbalanced}) under the Stirling approximation.

It is interesting to notice the competition of two behaviors at the edges $|x|=1$: when ${N\over4}(1+x)+{1\over2}$ is an integer the distribution vanishes, while when ${N\over4}(1+x)$ is an integer the distribution diverges like $(1-x^2)^{-{1\over4}}$. This is reminescent of the balanced input case with $y=0$ [see comments after (\ref{balanced})].

\section{Comments on the imbalanced-input case}
\label{sec:comments}

Starting from the approximate formula (\ref{13}), the average between the two slowly oscillating envelope curves can be estimated to be given by the function $P(x)$ in (\ref{eqn:Pave}), for any $Ny^2\lesssim1$. In this sense, the function $P(x)$ appears to be ``universal,'' in this context. Let us elaborate on this idea.

Let the initial input state be randomly picked up among states with input imbalance $Ny$ with equal probability. Assume that the input imbalance is bounded by a parameter $n = o(N)$, that is, $|y|\le n/ N \ll1$ for large $N$. Then the average distribution reads
\begin{equation}
{1\over n +1}\sum_{-n\le Ny\le n}{N\over4} |A_N(x,y)|^2 \equiv\overline{P_N(x)},
\label{avi}
\end{equation}
where the summation is taken over $n+1$ even values of $Ny$ (and $n$ is assumed to be an even number, for simplicity). In the sub-Poissonian case $n=o(\sqrt{N})$ we can disregard the exponential factor $e^{-{N\over4}y^2}$ arising from the prefactor in (\ref{13}) and take the average of the following quantities ($\phi={\pi\over2}+\sin^{-1}x$)
\begin{align}
{1\over n+1}\sum_{k=-{n\over2}}^{n\over2}{\sin^2|k|\phi\over(1-x^2)}&={1\over2(1-x^2)}\left(1-{\sin[(n+1)\phi]\over(n+1)\sin\phi}\right),\nonumber\\
{1\over n+1}\sum_{k=-{n\over2}}^{n\over2}\cos^2|k|\phi&={1\over2}\left(1+{\sin[(n+1)\phi]\over(n+1)\sin\phi}\right).
\label{averages}
\end{align}
Plugging these results in (\ref{avi}) one gets
\begin{widetext}
\begin{align}
\overline{P_N(x)}={1\over\pi} \Biggl[& {1\over2}\left(1-{\sin [(n+1)({\pi\over2}+\sin^{-1}x)]\over(n+1)\sqrt{1-x^2}}\right){1\over\sqrt{1-x^2}}\biggr\vert_{{N\over4}(1+x)+{1\over2}={\rm integer}}\nonumber\\
&{}+{1\over2}\left(1+{\sin[(n+1)({\pi\over2}+\sin^{-1}x)]\over(n+1)\sqrt{1-x^2}}\right){1\over\sqrt{1-x^2}}\biggr\vert_{{N\over4}(1+x)={\rm integer}}\Biggr].
\label{exactimb}
\end{align}
\end{widetext}
This is our third and last main result. We see that the oscillating behavior appearing alternatively at $Nx=0,\pm4,\pm8,\ldots$ and at $Nx=\pm2,\pm6,\ldots$ is canceled if we look at the average distribution (or more practically, if we are unable to distinguish the number states $|m_a, m_b\rangle$ and $|m_a\pm \delta m,m_b\mp \delta m \rangle$ at the output ports), which can be viewed as a universal quantity
\begin{equation}
\overline{P_N(x)}\Bigr\vert_{\rm typical}={1\over\pi}{1\over\sqrt{1-x^2}} = P(x),
\label{pbar}
\end{equation}
where $Nx$ is an even number.
The amplitude of the oscillations in $\overline{P_N(x)}$ vanishes as $1/n$ for large input imbalance $n$. This results is still valid in the Poissonian case, when $n=O(\sqrt{N})$: in such a case, the exponential factor $e^{-{N\over4}y^2}$ must be included and the average procedure can be conducted through Gaussian integrations. 

\section{Two-body correlation of the probability distribution (statistical fluctuations)}
\label{sec:twopoints}

The quantity $P(x)$ in (\ref{pbar}) is a common feature of all output distributions, being robust against the imbalance in the input photon numbers (the upper tolerable imbalance being Poissonian).
It is then interesting to study the effect of statistical fluctuations.

Consider a physical quantity $f(x)$ that is a function of the output imbalance $x$.
Such a quantity can be the $x$-representation of an operator $\cal O$, $f(x)=\langle x{|\cal O}|x\rangle$.  
Its statistical properties are governed by the variance of its expectation value over the probability distribution $P_N(x)$ and over the input imbalance $y$,
\begin{widetext}
\begin{equation}
\delta^2f(x)=\overline{\langle f^2(x)\rangle}-\overline{\langle f(x)\rangle}^2=\int dx\,dx'f(x)f(x')\,\Bigl(\overline{P_N(x)P_N(x')}-\overline{P_N(x)}\cdot\overline{P_N(x')}\Bigr),
\end{equation}
where $\langle f(x)\rangle=\int dx\,f(x)P_N(x)$ and the average $\overline{{}\cdots{}\vphantom{|}}$ over $y$ is defined in (\ref{avi}).
The terms in brackets represent the correlation function of the probability distribution, and are not difficult to evaluate, for the averages over $y$ can be calculated by explicitly summing up all possible integers $Ny$, like in (\ref{averages}). The result is
\begin{align}
&\overline{P_N(x)P_N(x')}-\overline{P_N(x)}\cdot\overline{P_N(x')}\nonumber\\
&={1\over\pi^2\sqrt{(1-x^2)(1-x'^2)}}{\epsilon(x,x')\over8(n+1)}\left({\sin[(n+1)(\phi+\phi')]\over\sin(\phi+\phi')}+{\sin[(n+1)(\phi-\phi')]\over\sin(\phi-\phi')}-{2\over n+1}{\sin[(n+1)\phi] \over \sin\phi}{\sin[(n+1)\phi']\over\sin\phi'}\right),
\end{align}
where $\phi={\pi\over2}+\sin^{-1}x$, $\phi'={\pi\over2}+\sin^{-1}x'$, and 
\begin{equation}
\epsilon(x,x')=
\begin{cases}
+1&\hbox{\rm both ${N\over4}(1+x)$ and ${N\over4}(1+x')$ are integers or both ${N\over4}(1+x)+{1\over2}$ and ${N\over4}(1+x')+{1\over2}$ are integers},\\
-1&\hbox{\rm both ${N\over4}(1+x)$ and ${N\over4}(1+x')+{1\over2}$ are integers or both ${N\over4}(1+x)+{1\over2}$ and ${N\over4}(1+x')$ are integers},\\
0&\hbox{otherwise}.
\end{cases}
\end{equation}
\end{widetext}
The range of input imbalance fluctuations $-n \le Ny\le n$ is assumed here to extend to a sub-Poissonian region $n=o(\sqrt{N})$.
Therefore, for large $n$, the above correlation function decays at most like $1/n$, realizing a ``typical" behavior $\delta f(x)\to0$.                 
Clearly, if one is unable to count the exact number of photons at the output ports, then the relevant probability distribution is given by the average (\ref{pbar}), that has lost the $y$ dependence, and thus no correlation survives.

\section{Concluding remarks}
\label{sec:concl}
We investigated the photon distribution at the output of a beam splitter for balanced and imbalanced input states.
Equations  (\ref{imbalanced})--(\ref{13}) and (\ref{exactimb})--(\ref{pbar})
generalize the Hong--Ou--Mandel scheme, according to which two identical photons that illuminate a balanced beam splitter always leave through the same exit port. 
In the limit of large $N$, the output distribution follows a $(1-x^2)^{-1/2}$ law, and the output state can be viewed as a generalized NOON state, as photons tend to appear at only one of the output ports.
We have seen that such an output distribution is robust and reminiscent of typical statistical behavior.

Our results are linked to the results obtained in Refs.\ \cite{Konno1,Konno2}: a beam splitter Hamiltonian implements a continuous-time quantum walk describing perfect state transfer in spin chains~\cite{MS}. This fact allows one to directly apply them also to spin dynamics under the exchange interaction. In the context of the recent research in multi-particle multi-mode quantum walks, it would be very interesting to extend our results to the case of multi-mode interferometers and mixed Fock input states.

\begin{acknowledgments}
We would like to thank the organizers of the conference ``Advances in Foundations of Quantum Mechanics and Quantum Information with Atoms and Photons" (INRIM, Turin, 2014) for giving us the opportunity to discuss the preliminary ideas on which this work is based, and Francesco Pepe for insightful remarks.
This work was supported by the Top Global University Project from the Ministry of Education, Culture, Sports, Science and Technology (MEXT), Japan.
S.P. was partially supported by the PRIN Grant No.\ 2010LLKJBX on ``Collective quantum phenomena: from strongly correlated systems to quantum simulators.''
M.S. was supported by the EU 7FP Marie Curie Career Integration Grant No.\ 322150 ``QCAT,'' by the NCN Grant No.\ 2012/04/M/ST2/00789, by the MNiSW Co-Financed International Project No.\ 2586/7.PR/2012/2, and by the MNiSW Iuventus Plus Project No.\ IP 2014 044873.
K.Y. was supported by a Grant-in-Aid for Scientific Research (C) (No.\ 26400406) from Japan Society for the Promotion of Science (JSPS) and by the Waseda University Grant for Special Research Projects (No.\ 2015K-202).
\end{acknowledgments}


\end{document}